\documentclass[sigconf, authorversion=true]{acmart}
\usepackage[english]{babel}
\usepackage[utf8]{inputenc}
\usepackage[T1]{fontenc}
\usepackage{setspace}

\usepackage{flushend}

\usepackage{enumitem}
\setlist[itemize]{leftmargin=1em}
\setlist[enumerate]{leftmargin=4ex}

\usepackage{hyperref} 
\hypersetup{
    pdfauthor={Sascha El-Sharkawy, Dhar Saura Jyoti, Adam Krafczyk, Slawomir Duszynski, Tobias Beichter, and Klaus Schmid},
    pdftitle={Reverse Engineering Variability in an Industrial Product Line: Observations and Lessons Learned},
    pdfsubject={22st International Conference on Software Product Line},
    pdfkeywords={Software product lines; variability modeling; reverse engineering; static analysis},
    pdfpagemode={UseOutlines},
    pdfproducer={LaTeX},
    colorlinks=true,
    linkcolor=black,          
    citecolor=black,          
    filecolor=black,          
    urlcolor=black            
}

\usepackage{framed}
\usepackage{xcolor}

\newcommand{\ifDef}{\texttt{\#ifdef}}
\newcommand{\ifNDef}{\texttt{\#ifndef}}
\newcommand{\errorDef}{\texttt{\#error}}
\newcommand{\ifDefBlock}{\ifDef-block}

\colorlet{shadecolor}{gray!25}
\newcommand{\Observation}[1]{
    \begin{snugshade*}
      \noindent \textbf{OBSERVATION}\\[0.5ex]#1
    \end{snugshade*}
}

\usepackage{listings}
\definecolor{cEclipsePurple}{rgb}{0.5,0,0.35}
\lstdefinelanguage{cWithPre} {
  language=C,
  moredelim = [l][\bfseries\color{cEclipsePurple}]{\#}
}

\usepackage{ifthen}
\newboolean{showToDos}
\setboolean{showToDos}{false} 
\ifshowToDos
  \usepackage{marginnote} 
  \let\marginpar\marginnote
  
  \newcommand{\ToDo}[1]{\noindent\fcolorbox{black}{yellow!20!white}{\parbox{.975\columnwidth}{#1}}}
  \newcommand{\SubjectedToDo}[2]{\ToDo{\textbf{#1:}#2}}
  
  \newcommand{\Outline}[1]{\color{blue}{#1}\color{black}}
  \newcommand{\Problem}[1]{\color{red}{#1}\color{black}}
  \newcommand{\secSize}[1]{[#1]}
  
  \usepackage[textsize=small,colorinlistoftodos]{todonotes}
  \newcommand{\userComment}[2]{\todo[color=#1!20,linecolor=#1!60,bordercolor=#1!60]{#2}}
  \newcommand{\ks}[1]{\userComment{orange}{\textbf{KS:} #1}}
  \newcommand{\se}[1]{\userComment{red}{\textbf{SE:} #1}}
  \newcommand{\ak}[1]{\userComment{blue}{\textbf{AK:} #1}}
  \newcommand{\sd}[1]{\userComment{green}{\textbf{SD:} #1}}

  \usepackage{pifont}
  
\else
  \newcommand{\ToDo}[1]{}
  \newcommand{\SubjectedToDo}[2]{}
  \newcommand{\Outline}[1]{}
  \newcommand{\Problem}[1]{}
  \newcommand{\secSize}[1]{}
  
  \newcommand{\ks}[1]{}
  \newcommand{\se}[1]{}
  \newcommand{\ak}[1]{}
  \newcommand{\sd}[1]{}
\fi

\begin{document}
\title[Reverse Engineering Variability in an Industrial Product Line]{Reverse Engineering Variability in an Industrial Product Line: Observations and Lessons Learned}


\author[S.\ El-Sharkawy]{Sascha El-Sharkawy}
\affiliation{
  \institution{University of Hildesheim}
  \streetaddress{Universitätsplatz 1}
  \city{31141 Hildesheim} 
  \state{Germany} 
}
\email{elscha@sse.uni-hildesheim.de}

\author[S.\ Dhar]{Saura Jyoti Dhar}
\affiliation{
  \institution{Robert Bosch GmbH}
  \streetaddress{Robert-Bosch-Straße 2}
  \city{71701 Schwieberdingen, Germany} 
}
\email{saura.jyoti@bosch.com}

\author[A.\ Krafczyk]{Adam Krafczyk}
\affiliation{
  \institution{University of Hildesheim}
  \streetaddress{Universitätsplatz 1}
  \city{31141 Hildesheim} 
  \state{Germany} 
}
\email{adam@sse.uni-hildesheim.de}

\author[S.\ Duszynski]{Slawomir Duszynski}
\affiliation{
  \institution{Robert Bosch GmbH}
  \streetaddress{Robert-Bosch-Straße 2}
  \city{71701 Schwieberdingen, Germany} 
}
\email{slawomir.duszynski@bosch.com}

\author[T.\ Beichter]{Tobias Beichter}
\affiliation{
  \institution{Robert Bosch GmbH}
  \streetaddress{Robert-Bosch-Straße 2}
  \city{71701 Schwieberdingen, Germany} 
}
\email{tobias.beichter@bosch.com}


\author[K.\ Schmid]{Klaus Schmid}
\affiliation{
  \institution{University of Hildesheim}
  \streetaddress{Universitätsplatz 1}
  \city{31141 Hildesheim} 
  \state{Germany} 
}
\email{schmid@sse.uni-hildesheim.de}



\begin{abstract}
Ideally, a variability model is a correct and complete representation of product line features and constraints among them.  Together with a mapping between features and code, this ensures that only valid products can be configured and derived.  However, in practice the modeled constraints might be neither complete nor correct, which causes problems in the configuration and product derivation phases.
This paper presents an approach to reverse engineer variability constraints from the implementation, and thus improve the correctness and completeness of variability models.

We extended the concept of feature effect analysis \cite{NadiBergerKastner+15} to extract variability constraints from code artifacts of the Bosch PS-EC large-scale product line. We present an industrial application of the approach and discuss its required modifications to handle non-Boolean variability and heterogeneous artifact types.


\end{abstract}

%
%





\setcopyright{acmlicensed}

\acmDOI{10.1145/3233027.3233047}

\acmISBN{978-1-4503-6464-5/18/09}

\acmConference[SPLC'18]{22nd International Systems and Software Product Line Conference}{September 10--14, 2018}{Gothenburg, Sweden}
\acmYear{2018}
\copyrightyear{2018}

\acmPrice{15.00}

\maketitle

\section{Introduction}
In software product line engineering, the accurate definition of constraints among elements of the variability model is not trivial to get right, but fundamental to defining the valid configurations. The constraints originate from domain knowledge, but may also reflect   dependencies in the implementation, e.g., which implementation components may work together. Consistent modeling of constraints requires much effort. However, this is hard to achieve or may be not a high priority. As a consequence, the variability model may deteriorate and require a correction at a later time. In this paper, we present observations and lessons learned while applying an automated approach to reverse engineer variability constraints from implementation artifacts of a successful and long-lasting industrial product line of the Robert Bosch GmbH.

The business unit Powertrain Solutions - Electronic Controls (PS-EC) of Robert Bosch GmbH\footnote{\url{http://splc.net/hall-of-fame/bosch/}} is a well-known Hall of Fame member with extensive experience in product line engineering. The development of an architecture with about 300 independently managed subsystems is one of the success factors of this product line and facilitates the delivery of about 2000 new product variants per year. The configuration process of this huge product line, while partially automated, involves the participation of functional subsystem experts providing their domain knowledge.  As part of an ongoing improvement of the development process, Bosch PS-EC aims to reduce the effort of configuring new product variants. To support that goal,  reverse engineering should be applied to gather as much as possible of the implicit domain knowledge from implementation artifacts.

In the reverse engineering activity, Bosch PS-EC pursues three objectives:
\begin{enumerate}[label=$\text{\textbf{O}}_\text{\textbf{\arabic*}}$]
	\item \label{obj:Lower Bound} In the future, the reverse engineered constraints should be incorporated into a variability model to guide the users while configuring new product variants. Based on a partial configuration, an automated configuration tool should hide features whose selection has become irrelevant. Thus, compliance with the reverse engineered constraints should not avoid any supported code configuration, i.e., the identified constraints should be a lower bound of the actual configuration space.
  \item \label{obj:Concise Constraints} The reverse engineered constraints should be as concise as possible. This is demanded as the constraints should provide to the developers a better understanding of the implemented dependencies and shall form the basis for the development of a new variability model.
  \item \label{obj:Obsolete variability} The results of the reverse engineering approach should be integrated with information received from other analyses to improve the quality and usefulness of the result. For example, Bosch PS-EC keeps track of legacy variability - features not used for configuration anymore, but kept to support legacy projects delivered in the past. This knowledge should be used to generate a variability model free of any legacy features.
\end{enumerate} 

\noindent As part of the ITEA3-project REVaMP2\footnote{\url{https://www.revamp2-project.eu/}}, we conducted a feature effect analysis \cite{NadiBergerKastner+15} to reverse engineer variability dependencies. This approach was evaluated on publicly available product lines like the Linux kernel. However, the development of the Bosch PS-EC product line differs fundamentally from the implementation of the studied open source systems, which requires extensions of the originally published feature effect analysis approach. These necessary modifications are presented in this paper.

Overall, we make the following contributions:
\begin{itemize}
	\item Experiences from industry regarding the reverse engineering of product line variability.
  \item Extension of the feature effect approach from \cite{NadiBergerKastner+15} to handle non-Boolean variability and heterogeneous artifact types.
\end{itemize}

\noindent The remainder of this paper is structured as follows. The next section describes the context of the presented study. Section~\ref{sec:Constraints Identification} presents our approach of reverse engineering variability dependencies of the Bosch PS-EC product line and points out important aspects, which need to be considered during the analysis. In Section~\ref{sec:Product-wise Analysis} we present the results of our analysis, before we discuss the usefulness for Bosch in Section~\ref{sec:Discussion}. Section~\ref{sec:Threats} reviews threats to validity and Section~\ref{sec:Related Work} presents related work. Finally, we conclude and outline future work in Section~\ref{sec:Conclusion}.

\begin{figure*}[!tbh]
	\centering
		\includegraphics[trim={3.5cm 6.75cm .75cm 4.48cm},clip,scale=.8]{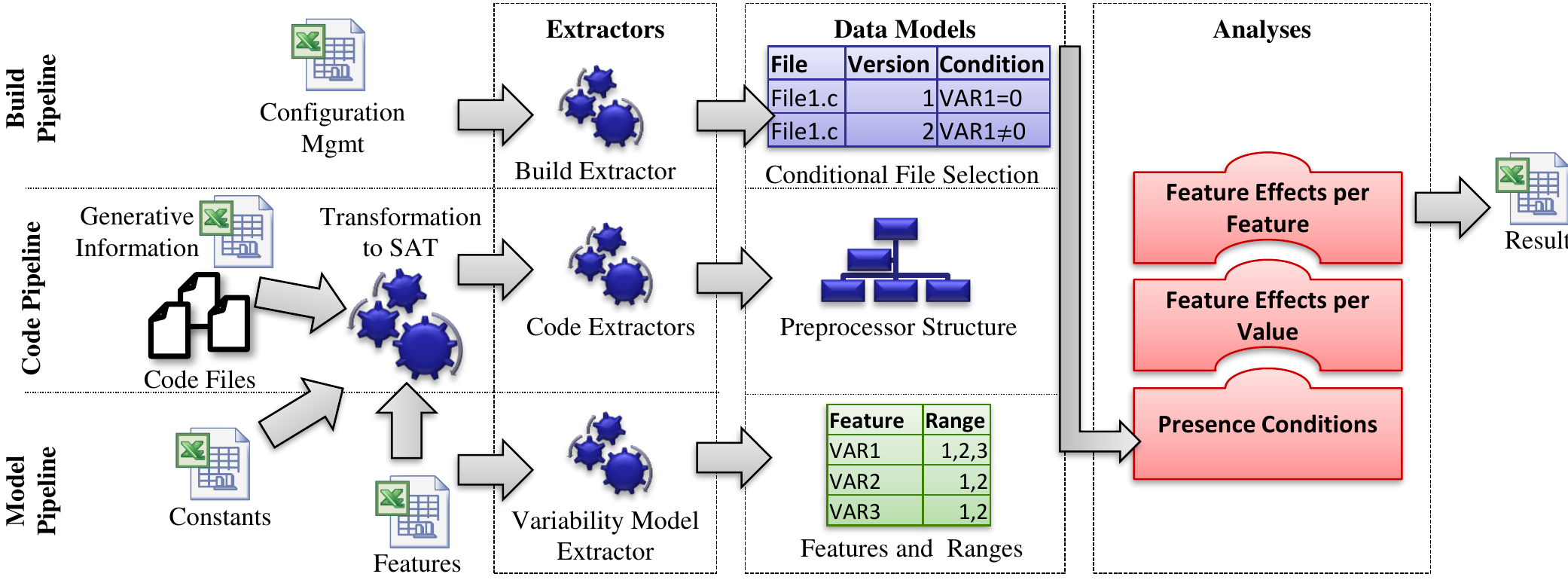}
	\vspace*{-2ex}
	\caption{Bosch-specific instantiation of the KernelHaven architecture for reverse engineering variability dependencies.}
	\label{fig:KH-Architecture}
\end{figure*}

\section{Context}
\label{sec:Context}

The business unit Powertrain Solutions - Electronic Controls (PS-EC) of Robert Bosch GmbH develops embedded electronic control units (ECU) for diesel, gasoline, hybrid, and electric engines. The application areas for the engines are diverse: from two-wheelers (scooters, motorbikes), through passenger cars, delivery vehicles and trucks, up to off-road applications such as construction machinery and stationary industrial engines. This involves supporting a~huge variety of product functionalities and an even bigger number of system and software variants to meet all the customer needs. The PS-EC engine control software is developed as a software product line, and was introduced in more detail by Tischer et al.\ in \cite{TischerBossMuller+12}.

\textbf{Variability drivers.} The variability of engine control software is driven by factors such as market-specific legislation, engine hardware variability, and divergent powertrain solutions (like hybrid vehicles, range extender concepts, electrically driven vehicles). Furthermore, customer-specific requirements concerning drivability, fuel economy, communication with other controllers in the vehicle and with the test equipment in the service center contribute to the variability. Variability also results from non-functional requirements: for example, different business models regarding organizational responsibility split are supported. These models range from Bosch acting as full service provider for the vehicle manufacturers, delivering the complete software and hardware, through various "Software Sharing" models involving a mixture of Bosch and customer software components, up to the delivery of only the ECU hardware with device drivers. On average, 2000 new  product variants are built and delivered to the customers every year. 

\textbf{Variability of the EC software.} The physical system controlled by the engine control software is very complex, with multiple sensors, control algorithms and actuators working together to create the most efficient chemical reaction in a real-time environment. Due to its size and complexity, the overall system is decomposed into numerous subsystems, each of which is defined and maintained by a team of experts on the particular technical domain. Hence, the subsystem configuration know-how is in general present as the domain knowledge of the function team. The product line consists of about 300 subsystems, from which about 100 subsystems are included in a typical product variant. The subsystem selection is different for each product variant: to configure the entire project, one must select the correct subsystem components and then configure the code level variations for each of them to get the proper software for the target hardware system. Hence, the function experts are closely linked to the process of defining, maintaining, and configuring variants for a particular project.

\textbf{Current concepts for managing variability.} The initial implementation concept of the product line was to have a single, all inclusive functional implementation of every subsystem. The variations were implemented within the code blocks guarded by preprocessor statements and were resolved during the preprocessing phase of the software build process. Unfortunately, this resulted in a significant complexity of the solution, causing a high effort for feature introduction, maintenance, and configuration. It also caused problems in parallel functionality development as different teams needed to work on the same code files.

This resulted in the diversification of the centralized product line strategy. The concept of maintaining code level variations using preprocessor statements is still retained. However, the single generic implementation is now split into several broad categories such as Platform, Customer Platform and Customer Projects where functionalities are  developed in parallel and maintained in separate files. Hence, the mutually related features of a subsystem are still developed in the same code files, but the complexity of that implementation is much lower because the other, not related features were moved to an alternative code module of the same subsystem. Consequently, the configuration of a subsystem now involves the selection of the suitable platform alternative, followed by setting the subset of preprocessor constants relevant for that alternative. Thus, the current approach treads the middle path between the variability mechanisms of file selection and conditional compilation.

Currently, the variability of EC software is well-managed based on the software architecture. The system decomposition is reflected in the variability space, as each feature and each preprocessor constant are clearly owned by a single subsystem or subsystem alternative. Additionally, each preprocessor constant, its dependencies, and possible values are documented, and changes to them are reviewed. To support correct configuration, further tool and process-based variation handling techniques were introduced. For example, the configuration management system is aware of preprocessor constants and uses their values in the component and branch selection. In the end, product variation can be maintained and configured with reasonable effort. However, managing variability still depends on the domain knowledge of the function experts.

\textbf{Re-introduction of variability modeling.} The main aim of the reverse engineering activity described in this paper is to support the creation of formal variability models. Global feature models existed in the product line in the past. However, the internationally distributed development of hundreds of components, existing in many production-relevant variants and versions and supporting hundreds of parallel project deliveries, made it impractical to maintain central variability models. In consequence, the global models became outdated and fell out of use. Presently a decentralized modeling strategy is introduced, with subsystem-level feature models maintained and versioned locally with the components, and assembled together for a product-level configuration.

At first, the subsystem variability models should be filled with reverse-engineered constraints. Afterwards, the constraints would be manually reviewed by the experts and further constraints, resulting from domain knowledge, would be added. The completed variability models could then be used for product configuration. Once the models are productive, it is expected that the dependency on function experts is significantly reduced, and a project configuration of identical quality can be achieved with lower effort.

\vspace*{-1ex}
\section{Constraints Identification}
\label{sec:Constraints Identification}
The feature effect analysis was developed to reverse engineer variability models based on code artifacts and build scripts and was validated on publicly available product lines like Linux \cite{NadiBergerKastner+15}. However, the authors limit their approach to study Boolean features only, which is sufficient for analysis of the studied systems, but differs fundamentally from the implementation of the PS-EC product line, which makes it necessary to adapt the approach to the needs of the particular product line. This section presents the extended approach to reverse engineer variability dependencies from the PS-EC product line, which is designed to resolve variability at development time (e.g., via preprocessor statements and module replacements). Below, we use the terms \textit{feature} to denote configurable elements of the problem space and \textit{variable} as their counter-part in the solution space, i.e., use in variation points of implementation artifacts. Further, we present the main differences as observations throughout this section.

First, we present the tool-chain, which we use to extract variability dependencies at Bosch PS-EC. The tool-chain is realized with KernelHaven \cite{KernelHaven}, which is also implemented as a product line in order to be customizable to conduct different analyses on product lines. Section~\ref{sec:Legacy Variability} presents a prior analysis to filter legacy variability in order to avoid dependency information of legacy features in the reverse engineered variability models. In Section~\ref{sec:Conceptual Foundation}, we explain the basic idea of feature effect analysis \cite{NadiBergerKastner+15}, which we used in this study. Afterwards, we show further extensions which are necessary for the given context. These are the translation of numerical expressions to propositional logic (cf.\ Section~\ref{sec:Mapping to SAT}), the simplification of the resulting formulas (cf.\ Section~\ref{sec:Concise Formulas}), and the analysis of heterogeneous artifact types (cf.\ Section~\ref{sec:Further Information Sources}).

\vspace*{-2ex}
\subsection{Tool for Reverse Engineering of Variability}
\label{sec:Technical Approach}
We use KernelHaven \cite{KernelHaven, KroeherEl-SharkawySchmid18, KroeherEl-SharkawySchmid18b} for the extraction of variability dependencies from implementation artifacts. Also KernelHaven is realized as a product line, which can be configured to run various analyses on product lines. Figure~\ref{fig:KH-Architecture} presents the instantiation of KernelHaven for reverse engineering the variability of the Bosch PS-EC product line.

KernelHaven provides three pipelines for the extraction of variability information from different variability spaces. Each of these extraction pipelines allows the flexible exchange of required extractor plug-ins to adapt KernelHaven to the needs of the analyzed product line. These extraction pipelines are:

  The \textbf{Build Pipeline} is responsible for the extraction of conditional inclusion information of code files during the build process. This conditional inclusion may be realized via \texttt{make}-scripts or, in case of the Bosch PS-EC product line, through a configuration management system providing the selection of alternative implementations of the same subsystem. This variability realization technique is sometimes also named as ``module replacement'' \cite{FussbergerZhangBecker17}.

  The \textbf{Code Pipeline} is responsible for the extraction of fine-grained variability information within code artifacts, e.g., preprocessor statements. We realized 4 different extractors for the extraction of preprocessor-based variability \cite{KernelHaven_wiki}. Each of these extractors can be flexibly interchanged and brings their own advantages and disadvantages. For the extraction of preprocessor variability in this case study, we use a block-based code extractor~\cite{KH_CodeBlockExtractor}, which runs very fast and provides good results without a complex configuration. Further, we use an additional extractor to extract variability from MSR files, which are used for the generation of conditional interfaces. This is described in more detail in Section~\ref{sec:Further Information Sources}.

  The \textbf{Model Pipeline} is intended to extract information from a variability model. For the analysis of Linux, this pipeline is used to convert the complex variability model into CNF representation. The purpose of this study is to reverse engineer dependency information from the implementation artifacts, for which no modeled constraints are available. However, we use this pipeline to extract feature definitions and their ranges from the configuration system using a custom extractor.

KernelHaven allows the optional execution of \textbf{preparators} before the extraction phase, if required. These preparators are intended to normalize analyzed artifacts in order to facilitate reuse of extractors even if the analyzed artifacts are in an unexpected format. In case of the Bosch PS-EC product line, one preparator is used to transform numerical expressions into propositional logic. This is explained in more detail in Section~\ref{sec:Mapping to SAT}.

KernelHaven allows the flexible definition of an analysis pipeline based on the extracted data (cf.~right part of Figure~\ref{fig:KH-Architecture}). The analysis pipeline is composed of analysis components, which may consume data from extractors or from other analysis components. This concept allows reusing existing analysis components rather easily. For reverse engineering the variability of the Bosch PS-EC product line, we decided to reuse components for the detection of presence conditions and for the computation of feature effects as described in Section~\ref{sec:Conceptual Foundation}. The third analysis component is than used to compute variability model constraints based on the previous feature effect analysis component. This component produces the desired results.

The architecture of KernelHaven allows the reuse of the produced results also in further analysis. In the presented study, the output of the feature effect analysis can be used as input to validate the results based on past configurations (not shown in the figure).

\vspace*{-1em}
\subsection{Influence of Legacy Variability}
\label{sec:Legacy Variability}
Bosch PS-EC tracks legacy features which are no longer used for new product configurations. Despite not being used, these features still need to be kept in specific versions of the development artifacts to support legacy projects delivered in the past. For example, there might be a need to apply a different feature configuration, including a legacy feature, to a software version developed several years ago. Therefore, the legacy features represent valuable assets which enable Bosch PS-EC to provide long-time product support. The legacy features are classified into categories such as features not used anymore, used only with a single allowed value, used only by a particular customer etc. The development infrastructure ensures that the legacy features cannot be reactivated for new product development. Newly emerging legacy features are identified by reverse engineering and analysis of development databases.

The newly created variability models should only include the non-legacy features. Because of that, also the KernelHaven analysis should account for the fact that a part of the input features might be not relevant or only be allowed to assume a single value. Although this situation is specific to Bosch PS-EC, a similar requirement might exist in other long-living industrial product lines.

In the KernelHaven input, we listed the legacy features having a single value as constants. Consequently, in the evaluation and simplification of feature effect formulas each legacy feature is replaced by the respective constant. To estimate the effect of including the legacy information, we analyzed an example product variant without and with using that information. Out of 100\% features obtained in the first analysis, not using the legacy information, 1.6\% were treated as legacy and removed in the second analysis. Furthermore, 0.3\% of the features which were dependent in the first analysis became independent, in addition to 67.2\% of features which were independent in both cases. Finally, for 0.7\% of the original features, the feature effect condition was simplified due to the replacement of legacy features by constant values. Note that the analyzed product variant consists of generic code: while the variability mechanisms of component and branch selection have been already applied during product checkout, the code has not been preprocessed yet. Thus, the product code contains so-called 120\% variability, as compared to the 150\% variability of the product line and 100\% variability of the completely configured product variant.

The relative amount of legacy features increases when more products are added to the analysis. This is because the non-legacy features are used in many products, while the legacy features mostly have a scope limited to a specific customer platform or component version. Hence, adding a new product to the analysis introduces new unique legacy features, while the non-legacy features repeat themselves. For the 7 products studied in Section~\ref{sec:Product-wise Analysis}, 3.6\% of unique features used in at least one product belong to the legacy group.

We conclude that it is important to consider the legacy information in the specific case of the engine control software. Furthermore, the measured reduction of the analysis scope indicates that although the amount of impacted features is not high, there is some potential to remove the compilation conditions referring to the legacy features.

\Observation{Consideration of legacy variability reduces the amount of false positives among the reported dependent features.}

\vspace{-1ex}
\subsection{Feature Effect Computation}
\label{sec:Conceptual Foundation}
Feature effects express under which condition a variable has an impact on  product derivation. A feature effect ($FE_{f}$) for a feature $f$ is a combination of all presence conditions, which make use of $f$. Based on \cite{NadiBergerKastner+15}, we use the term \textit{presence condition (PC)} to refer to a propositional expression over features that determines when a certain code artifact is compiled. This comprises very fine-grained \ifDefBlock{s} surrounding a single line in a C-file \cite{SinceroTartlerLohmann+10}, but also the definition of conditional compilation of complete C-files \cite{BergerSheLotufo+10, NadiHolt11, DietrichTartlerSchroderPreikschat+12}. For this reason, we use for the identification of feature effects the presence conditions from code files as well as from the build system.

In the case of the Bosch PS-EC product line, conditional inclusion of code files is not controlled by build scripts. Instead, the configuration management system checks out alternative versions of a subsystem based on a partial configuration. From the perspective of computing feature effects, there is no conceptual difference between conditional compilation of code files controlled via make scripts or a configuration management system. In both cases, the conditions for code file selection must be considered for the accurate computation of presence conditions of all code parts. In KernelHaven this is supported by a specialized build extractor for the PS-EC product line. 

For a feature $f$ and $PCs_{f}$, the presence conditions it is used in, the feature effect is defined as \cite{NadiBergerKastner+15}\footnote{$PC[f \leftarrow C]$ means replacing each occurrence of $f$ in $PC$ with constant $C$; $\oplus$ denotes the \texttt{XOR}-operation}:

\vspace*{-2ex}
\begin{equation}\label{eq:feature effect}
  FE_{f} \coloneqq \bigvee_{PC\ \in\ PCs_{f}} PC[f \leftarrow True] \oplus PC[f \leftarrow False]
\end{equation}
\vspace*{-2ex}

The feature effect analysis relies on propositional logic, which allows the use of SAT solvers to effectively query if preconditions of features are already violated with regard to the current configuration. On the other hand, this approach supports only the input of propositional formulas. This is required to be able to solve the exclusive-or relations of Formula~\ref{eq:feature effect}.

\begin{figure}[!tb]
	\centering
		\includegraphics[trim={0cm 8cm 14cm 0.25cm},width=\columnwidth]{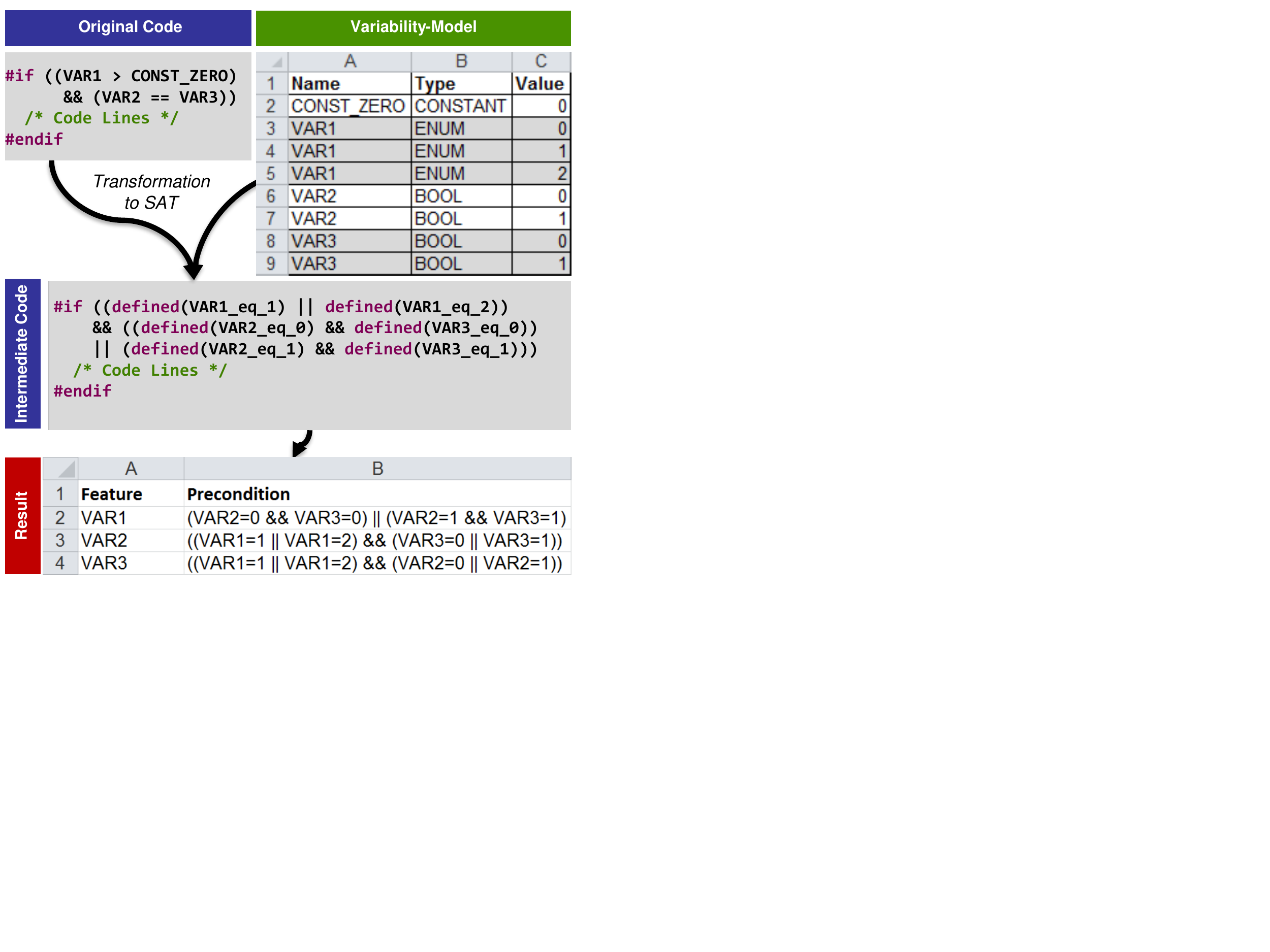}
	\caption{Transformation of numerical expressions into propositional logic.}
	\label{fig:Mapping to SAT}
	\vspace*{-4ex}
\end{figure}

\subsection{Mapping Numeric Expressions to SAT}
\label{sec:Mapping to SAT}
The original approach of computing feature effects \cite{NadiBergerKastner+15} was only defined for propositional logic, as presented in the previous section. This approach is sufficient for the analysis of the studied open source systems, because non-Boolean variability only plays a minor role in these systems. 
This does not apply for the implementation of the Bosch PS-EC product line, which uses expressions over numerical values very often. Most of the expressions are relational expressions ($=, \neq, >, \geq, <, \leq$), but also a few algebraic expressions are used (e.g., check if a certain bit is set or if the sum over 3 features is greater than a constant). However, most of the used features have a fixed range as they are handled either as Boolean variables (with range $[0, 1]$), enumerations, or constants. But the implementation contains also features representing integers and floats.

As the range of variables is typically very limited and the computations and comparisons are rather simple, we decided to transform the numerical expressions in the conditional compilation into propositional logic. This is shown in Figure~\ref{fig:Mapping to SAT}. Our approach allows us to reuse existing tooling, but requires two additional information sources for a correct translation as illustrated in the upper-right corner of the picture. 
\begin{itemize}
  \item First, we require a list of all features and their ranges of the Bosch PS-EC product line. This is used to transform equality and inequality expressions into define statements. We create for each valid variable-value pair a new Boolean variable, which represents the respective value assignment for the given variable. For features of type integer or float, we do not expand the whole possible range and translate them into a statement validating that the variable is defined at all. Hence, resulting feature effect conditions evaluate to true, if the integer/float is set to an arbitrary value. As a consequence, the resulting conditions are potentially more generalized (cf.\ \ref{obj:Lower Bound}), but are still useful to indicate dependencies between features.
	
  \item Second, we use a list of known constants (implementation-specific constants like debugging/physical constants, which shall not be part of the reverse engineered constraints) to simplify the analyzed expressions and increase the accuracy of the transformation. In the above example, \texttt{CONST\_\-ZERO} is replaced by 0 and the inequality expression can be mapped to valid values of \texttt{VAR1}.
\end{itemize}

\noindent This code transformation is done in a preprocessing step, before the code files are passed to the code extractors. The complete approach is explained in more detail in \cite{KrafczykEl-SharkawySchmid18}.

\begin{figure}[!tb]
	\centering
\begin{lstlisting}[language=cWithPre,caption={Preprocessor structure of Bosch PS-EC C-files.},label=lst:C-Example,basicstyle=\small]
#if ((VAR1 > 0) && (VAR2 != 0))
#if (VAR3 != 1)
/* Code Lines */                      %*\label{line:var3_code}*)
#endif
#else
/* Code Lines */
#endif

/**** test used features ****/ %*\label{line:separator}*)
#if ((VAR1 > 0) && (VAR2 != 0))       %*\label{line:check_begin}*)
#ifndef VAR3
#error >>>> 'VAR3' undefined!         %*\label{line:var3_error}*)
#endif
#endif %*\label{line:check_end}*)
\end{lstlisting}
		\vspace*{-5ex}
\end{figure}

Also, the information regarding the conditional checkout (and compilation) of C-files require a translation from numerical expressions into propositional logic. This information is passed via Excel spreadsheets to KernelHaven, which required the development of a specific build extractor. For this reason, we decided to integrate the translation algorithm directly into the extractor.

\Observation{Extraction of variability dependencies in the Bosch PS-EC product line requires analysis of numerical expressions.}

The code files may also contain consistency checks to ensure that variables are defined when used in preprocessor statements (cf.\ Lines~\ref{line:check_begin}~--~\ref{line:check_end} of Listing~\ref{lst:C-Example}). These checks make use of \errorDef-statements nested in \ifNDef-blocks to enforce the preprocessor to abort with an error, when a required variable definition is missing. The \ifNDef-statements change how the variable appears in the resulting presence conditions -- instead of a numeric expression, the variable is used in a Boolean negation (e.g., Line~\ref{line:var3_code} results in \texttt{VAR1 > 0 $\land$ VAR2 $\not=$ 0 $\land$ \color{green!50!black}VAR3 $\not=$ 1\color{black}}; Line~\ref{line:var3_error} results in \texttt{VAR1 > 0 $\land$ VAR2 $\not=$ 0 $\land$ \color{red!50!black}!VAR3\color{black}}). In contrast to the study of Nadi et al.\ \cite{NadiBergerKastner+15}, these consistency checks contain more coarse-grained variability information than the associated code parts. Another aspect is that these consistency checks are mostly auto generated by the development infrastructure, but may also be edited by developers and, thus, need not be correct. For these two reasons, we decided to omit these expressions from the analysis and remove them as part of the aforementioned preprocessing step of KernelHaven. However, we plan to use these expressions in a configuration mismatch analysis \cite{El-SharkawyKrafczykSchmid17} to verify our results.

\Observation{Extraction of variability dependencies needs to consider consistency checks (\errorDef-directives) as the resulting presence conditions are inconsistent with presence conditions of the remainder of the system.}

\vspace*{-1em}
\subsection{Concise Formulas}
\label{sec:Concise Formulas}
A major goal was that the resulting constraints should be easy to understand by human experts. Thus, the identified formulas need to be rewritten appropriately as the approach only provides rather complex and verbose constraint formulations 
(cf.~\ref{obj:Concise Constraints}).
In order to achieve this goal, we combined several simplification mechanisms. However, the use of these simplification mechanisms makes it harder to determine the relation to the original code dependence structure. For this reason, we provide capabilities to disable most of the simplification techniques to simplify the manual verification of the generated results.

Simplification of \textbf{Presence Conditions}. This step gathers all expressions to include a certain code fragment into the final product, which comprises the expansion of surrounding preprocessor conditions. Through the expansion, the resulting formulas may become very complex. KernelHaven supports the optional removal of all constant (sub-) expressions from propositional formulas, e.g., application of idempotence or absorption rule. This is realized via a third party library \cite{jBool}. Further, we omit duplicated formulas since they do not affect the feature effect generation. Due to the computational complexity, we compare only whether formulas are syntactically equal.

Simplification of \textbf{Feature Effects}. This analysis computes feature effects as explained in Section~\ref{sec:Conceptual Foundation}, based on the previously gathered presence conditions. This approach offers several potentials for further simplification mechanisms, which are supported by KernelHaven:
\begin{enumerate}
  \item During the computation of the xor-literals ($PC[f \leftarrow True] \oplus PC[f \leftarrow False]$), we resolve all constant expressions to keep the outcome as concise as possible.
  \item Finally, we apply the same simplification rules for propositional logic to the computed feature effects (disjunction over all xor-literals) as already done for the presence conditions. In earlier steps only the distinct literals are simplified. This step ensures simplification of complex results, if the disjunction over multiple presence conditions form new potential for simplification.
\end{enumerate}

\subsection{Further Information Sources}
\label{sec:Further Information Sources}
Above, we explained how we handle code files and build information to compute feature effects. However, the Bosch PS-EC product line has further artifacts containing variability, which are used for the derivation of products. Also, this information needs to be considered in order to compute correct feature effects. 
For this, we developed a new extractor to handle artifacts specific to the Bosch PS-EC product line. Below, we give a brief overview of the additional information sources and how they are considered.

\textbf{Interface descriptions.} In addition to code files, the Bosch PS-EC product line provides capability to generate interfaces according to selected features. This information is stored in XML files following the MSR, MDX, and AUTOSAR standard formats. The information in these files are used to activate/deactivate selected code blocks when the respective subsystem is selected by the configuration management system. The development infrastructure of Bosch provides capabilities to export this information. We incorporate the information of extracted interface variability with the variability information from code files (cf.\ code pipeline of Figure~\ref{fig:KH-Architecture}). When computing presence conditions, we need to consider the conditions of the configuration management system about the selection of the MSR files as well as the contained variability information of the MSR exported file.

The integration of \textbf{further artifact types} is planned as future work. For instance, this may include variability information of Perl scripts, which are used to generate code artifacts. The selection of the Perl scripts is  handled in a similar way as the selection of code files or the interface descriptions. These scripts contain variability information, which is resolved when generating code artifacts.

\Observation{Complete extraction of variability dependencies requires adequate interpretation of various artifact types.}

\section{Analysis of product variants}
\label{sec:Product-wise Analysis}

In contrast to the open source product lines analyzed by Nadi et al.\ \cite{NadiBergerKastner+15}, the code base of the Bosch PS-EC product line cannot be analyzed in its entirety. As discussed in Section~\ref{sec:Context}, the product line uses component selection, branching, and conditional compilation to instantiate pre-build variability. Furthermore, there exist dependencies between the different versions (i.e., revisions in time) of the cooperating components, so that many active products do not use the latest component versions. Hence, determining the component branches and versions which should be considered as the active code base depends on the selection of relevant product variants, and not on the properties stored in the code and the configuration management system alone. For our analysis we selected 7 recent product variant projects delivered to the key customers and using two different engine types (gasoline and diesel). The selection aimed at a possibly diverse product set, including both the base platform and the customer platforms code. As noted in Section~\ref{sec:Legacy Variability}, the code analyzed for a specific product contains so-called 120\% variability, as the variability mechanisms of component and branch selection have been already applied during product checkout, while the code has not been preprocessed yet.

\begin{figure}[!tb]
	\centering
		\includegraphics[trim={0.1cm 17.2cm 15.7cm 0.1cm},width=0.85\columnwidth]{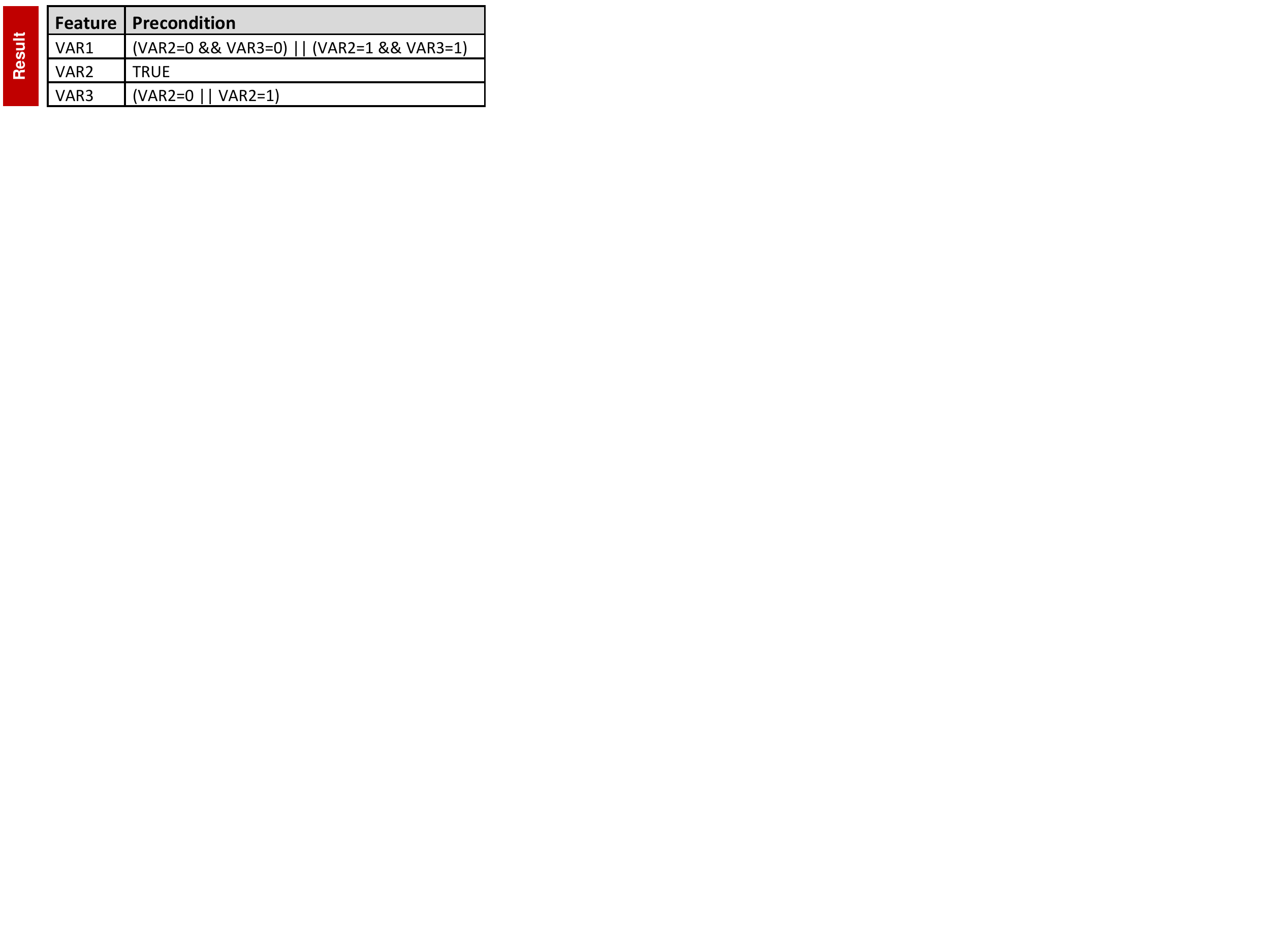}
	\caption{Final analysis result, if \texttt{VAR2} and \texttt{VAR3} are used in further variation points than shown in Figure~\ref{fig:Mapping to SAT}.}
	\label{fig:Final Result}
	\vspace*{-4ex}
\end{figure}

The primary analysis focus was the identification of all the features and their dependencies which are present in the product variant projects and also to determine whether a given feature represents a top level feature, which selection is not guarded by a feature effect, or a dependent feature, which use is limited by a feature effect. As discussed in the previous section, the focus is on variation points as they represent a feature inside code files. Based on the evaluation of the logical expression guarding a variation point, the underlying code block representing a particular feature gets activated or deactivated. The nested logical expressions help to determine the dependency of a particular feature on the availability or unavailability of another feature. This is illustrated in Figure~\ref{fig:Final Result}. \texttt{VAR2} is an independent feature, whose selection always affects the code (precondition \texttt{true}), while \texttt{VAR1} and \texttt{VAR3} are dependent on the selection of other features.
 
\begin{figure}[!b]
     \centering
     \vspace{-1em}
           \includegraphics[trim={0.1cm 5.25cm 0.6cm 0.8cm}, width=\columnwidth]{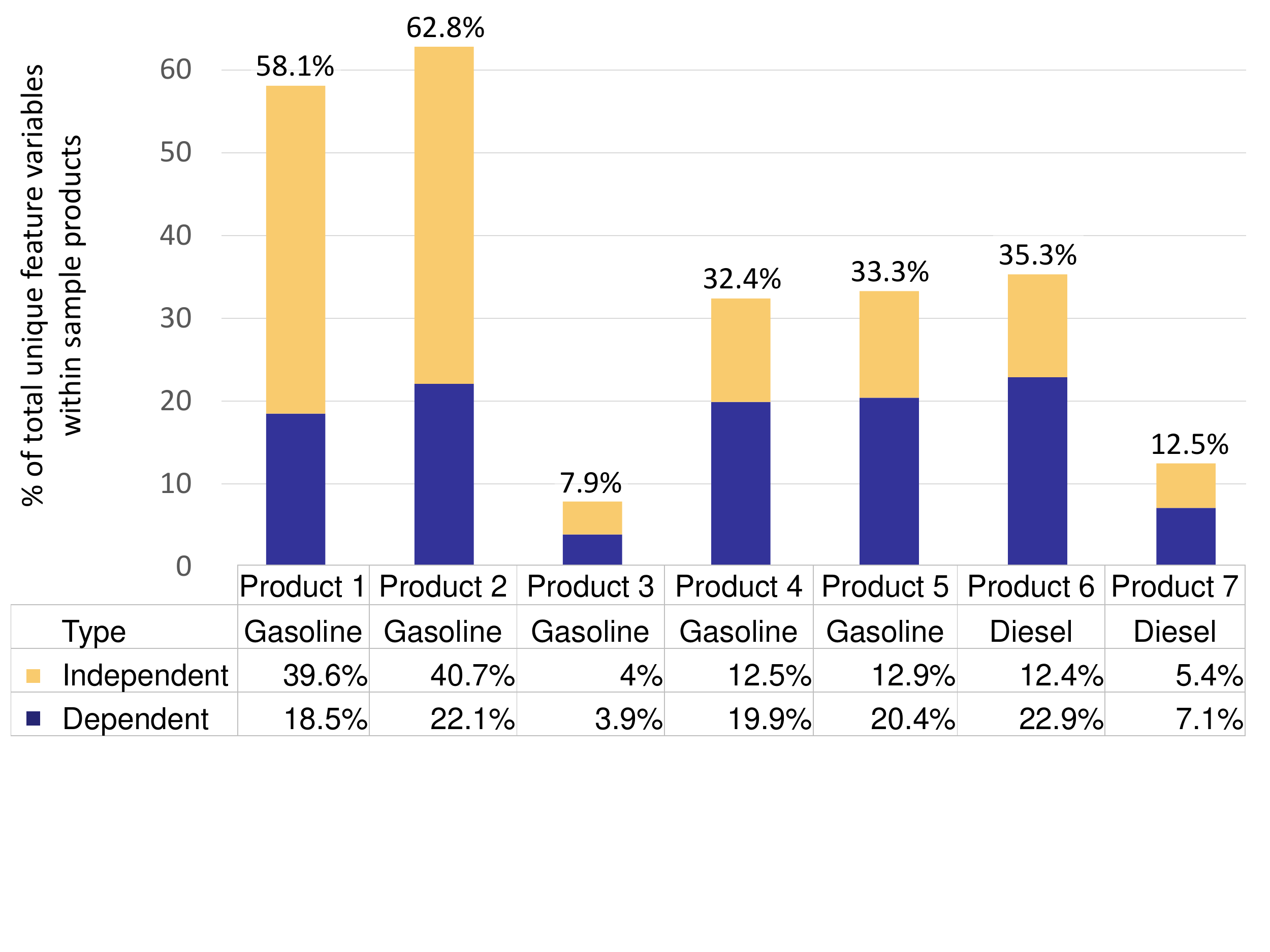}
     \caption{Feature dependency distribution across sample product projects.}
     \label{fig:Dependency distribution}
     \vspace{-1em}
\end{figure}

\textbf{Product-level analysis results.} Using KernelHaven, we ran the analysis for 7 product variant projects. For each product, we obtained the feature variables which were referred to in the code as well as the respective feature effect. We categorized the feature variables into two groups: independent ones, which had no feature effect (and hence should be considered to be the top level of the feature hierarchy), and the dependent ones for which a feature effect existed. The results of this analysis are depicted in Figure~\ref{fig:Dependency distribution}. 

Please note that for confidentiality reasons the analysis result does not specify the absolute numbers of feature variables in each product. Instead, the total number of unique feature variables used in any of the 7 products was set as the reference number, equal to 100\% of analyzed feature variables. Hence, a given variable name was counted once, regardless of the amount of products in which it occurred. Subsequently, each of the other numbers (e.g., amount of features of a given category in a given product) was divided by the reference number, obtaining the percentage values depicted in Figure~\ref{fig:Dependency distribution}. Consequently, all the size proportions of the reported feature variable groups are kept (e.g., Product~2 uses about 8 times as much feature variables as Product~3), but the absolute numbers of the variables are hidden. 

The relative proportion of dependent features, i.e., the number of dependent features in a product divided by the total number of features in that product, varies from 31.8\% in Product~1 up to 64.9\% in Product~6. This indicates that the code of analyzed products differs in the amount of used preprocessor expressions. We consider these results to show a high potential to reduce the effort of configuring new product variants. Furthermore, the structure of the reverse engineered constraints indicates that a graph-based structure is more appropriate to visualize the feature dependencies than a FODA-like tree structure.

\textbf{Result aggregation across multiple products.}
In the next step, we aggregated the results obtained for the individual products to derive a set of constraints valid for all analyzed products, which could subsequently be used for variability modeling. For each feature variable, we checked its feature effects obtained from all the analyzed products. If the feature effect was consistent, i.e., the feature had the same category for all products in which it occurred, we retained the categorization of that feature. For the opposite case, when different feature categories were obtained for some products, we created a third group of mixed feature variables. Thus, a mixed feature variable can be independent in one product and dependent in another. We consider that the feature effects of feature variables from the mixed group cannot be used as a basis for variability modeling, because they are not fulfilled by some of the products.

\begin{figure}[!tb]
	\centering
		\includegraphics[trim={0cm 3.75cm 0cm 0cm},width=.95\columnwidth]{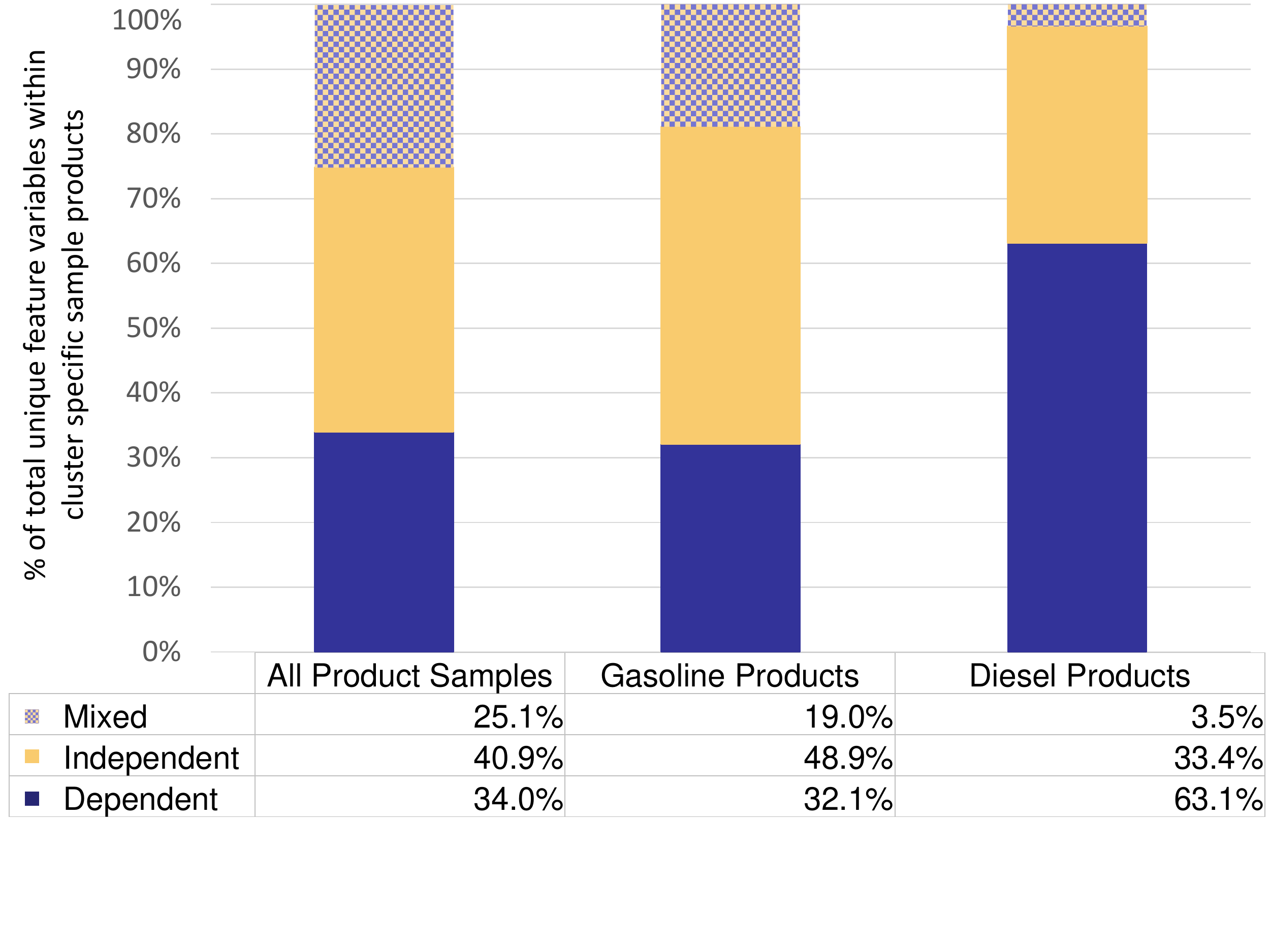}
	\caption{Feature dependency distribution aggregated upon product clusters.}
	\label{fig:Dependency distribution upon clusters}
	\vspace{-4ex}
\end{figure}

Figure~\ref{fig:Dependency distribution upon clusters} depicts the aggregation results across all 7 products, as well as across the diesel group (2 products, using 37.9\% of all feature variables) and the gasoline group (5 products, using 93.5\% of all feature variables). In the aggregation across all products, we observe that 25.1\% of all feature variables are classified as mixed -- this means that the number of feature variables with consistent feature effects decreased rapidly with the analysis of additional products from the product line. We interpret this as an effect of significant differences in the code basis of the products, due to the use of component selection variability mechanism. However, these differences should be lower across similar products, e.g., such using the same engine technology. The aggregations of both diesel and gasoline products suit this assumption, as they contain respectively 3.5\% and 19.0\% of mixed feature variables. At the same time, the aggregation of the diesel products contains a higher relative percentage of the dependent features (63.2\%) than the aggregation of all 7 products (34.0\%), while for gasoline that percentage is only marginally lower (32.1\%). Hence, the results obtained for product clusters indicate a higher potential than the results obtained for the unclustered product line.\looseness-1

\textbf{Reliable dependencies for many products.}
For an increasing number of analyzed projects, the aggregation of feature effects over these products is likely to contain an increasing proportion of feature variables belonging to the mixed group. This is due to the fact that a feature effect found to be consistent in a newly analyzed product does not change the categorization of a feature variable, while a found inconsistency changes the categorization to the mixed one. At the same time, feature effects of dependent variables are more valuable if they are found in many product, and not just one. To evaluate the effect of aggregating the analysis for many products, and to categorize the found consistent dependencies, we grouped all feature variables according to the number of analyzed products where they occur. 

\begin{figure}[!tb]
     \centering
           \includegraphics[trim={0cm 5cm 0cm 0cm},width=.985\columnwidth]{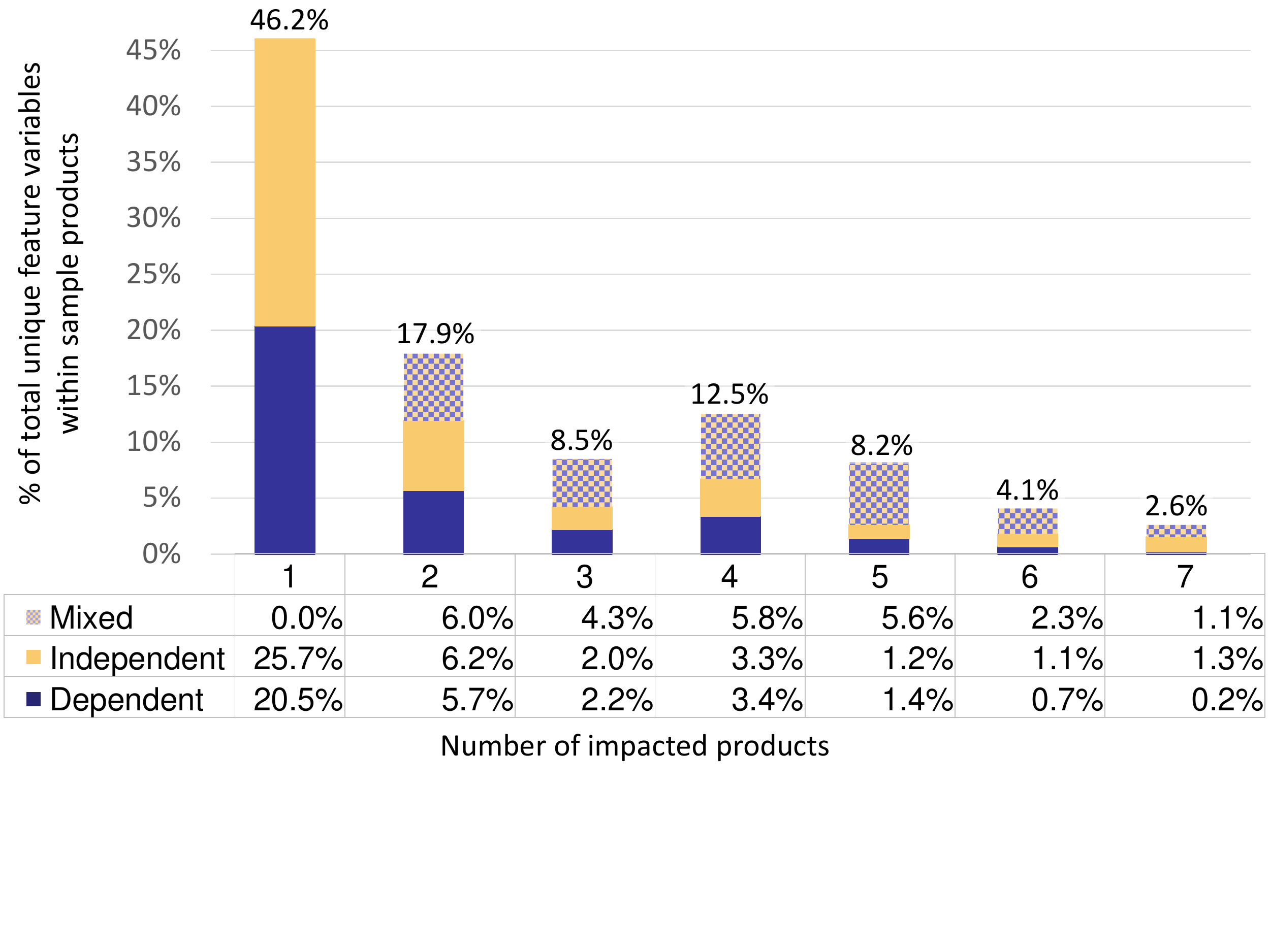}
     \caption{Feature dependency distribution aggregated upon number of source products.}
     \label{fig:Dependency distribution upon projects}
     \vspace{-4ex}
\end{figure}

The result of that analysis is depicted in Figure~\ref{fig:Dependency distribution upon projects}. 46.2\% of the analyzed feature variables are found in only one product. 20.5\%~of feature variables are such which are dependent and found in exactly one product. As no contradiction can be identified for just one constraint, there are no mixed category features used in exactly one product. Interestingly, all the other dependent feature variables, found in exactly 2 up to exactly 7 products, make up only 13.6\% of all analyzed feature variables -- compared to 25.1\% of all analyzed variables which are mixed in these groups. Note that as each feature variable is categorized to exactly one group (e.g., independent variables found in exactly 3 products), the percentages provided in Figure~\ref{fig:Dependency distribution upon projects} sum to 100\%.

In the above analysis, the chance of finding a consistent dependency in two or more aggregated products is low: the group of 38.7\% variables belonging to more than 1 product and having some dependency is split between 13.6\% dependent variables and 25.1\% mixed variables. We attribute it to the fact that the products selected for the analysis were very diverse -- which is also exemplified by the amount of features which were found in one product only.

\vspace*{-1em}
\section{Discussion}
\label{sec:Discussion}
The approach we describe in this paper is based on the current status of the Bosch activities to extract variability information based on existing data. The first goal of Bosch PS-EC was to \textbf{save configuration effort} by analyzing the hierarchies of variation points within the engine control software. As first results showed that only a part of the hierarchy is represented in the source code, additional variability dependencies from other sources such as configuration management were added as input for KernelHaven. This improved the result, allowing us to identify the features which depend on the configuration of other related features. These dependent features do not need to be configured if the related features were deactivated before. This activity holds significance as Bosch product line contains a significant amount of non-Boolean features offering multiple variations of the same feature type. The configuration of such features requires more time than Boolean features. The identification of feature hierarchies would lead to a pre-selection/elimination of the possible options offered by these non-Boolean feature variables, 
thereby reducing feature configuration effort.

The identified dependencies, hierarchies and affected system constants give a valuable input for \textbf{variability modeling}, serving as a starting point for new variant models. A developer gets information concerning existing dependencies within the source code and can use this to model the dependencies between modeled features and feature variables. The developer can also use the dependencies and hierarchies to \textbf{validate existing variant models} to ensure that they fit to the source code and are up-to-date. In both cases, the final benefit for Bosch is the time saved for setting and reviewing of the configuration. Further reduction potential is expected concerning saving test effort through a better use of the feature dependency information during test case definition.

Bosch PS-EC has a huge number of products within its product line. In REVaMP2, we focus on the extraction of variability models based on product variant projects. As the variants were developed for different customers, are based on different requirements, and are used for different automotive systems (e.g., gasoline/diesel, passenger car/commercial vehicle), the resulting low similarity of products might lead to a low amount of recoverable hierarchy information. By grouping similar sets of products for extraction, \textbf{larger product set-specific hierarchies} can be obtained, as exemplified by the evaluation data in the previous section.

The KernelHaven tool supports a \textbf{high flexibility} in extracting variability data from different sources like source code and proprietary Bosch PS-EC file formats. The tool uses several input data files, which we could fill by exporting the relevant data from development databases. This eliminated the need for writing database connectors and enabled the fast creation of the first results. This flexibility enables KernelHaven to be applied in different development contexts with different input sources.

As a summarized result, the ability to extract variability information with KernelHaven gives Bosch PS-EC new opportunities for saving effort and increasing quality in variant management. Further investigations will define what finally will be used in production.

\vspace*{-1em}
\section{Threats to Validity}
\label{sec:Threats}
In this section, we discuss threats to validity regarding the reverse engineering of variability dependencies reported above.

\textbf{Internal validity.} This aspect stresses to which extent the reverse engineered variability dependencies are correctly recovered from the artifacts, and are not too restrictive (cf.\ \ref{obj:Lower Bound}).

\textit{Correctness of code preparation.} In Section~\ref{sec:Mapping to SAT} we discussed the importance of preprocessing the code to handle numerical expressions and also to remove consistency checks. The exported range specifications of the used features allow to enumerate all possible value selections and, thus, the translation into propositional logic. For some features  no exported range definition is available. This conceptually affects the accuracy of the presented results as we compute only constraints regarding the selection of unbounded features rather than for individual value selections. Second, to avoid false computation of feature effects, we remove all consistency checks from the code files. For this, we do not remove only auto-generated blocks at the end of files, but also consistency checks manually added by developers at arbitrary positions. For this we detect all preprocessor blocks containing only \errorDef-statements and remove the complete block.\looseness=-1

\textit{Accuracy.} The accuracy of feature effect computation is influenced by the complete extraction of all preprocessor blocks from code files, but also from the correct preparation of code files, which was discussed above. The extraction of expressions from code files is currently done by a block-based code extractor \cite{KH_CodeBlockExtractor}. This tool is very fast and does not require a detailed configuration, like for example TypeChef \cite{KastnerGiarrussoRendel11, TypeChef}. However, the used extractor does not provide a sound and complete parsing algorithm for presence conditions, e.g., it neither expands macros, nor does it provide type-checking capabilities. This bears the risk of missing conditions, which further restrict the configuration space and, thus, are important to meet \ref{obj:Lower Bound}. This may only happen if \textit{all} instances of an incorrectly computed presence condition are actually more complex than detected, which can happen under rather special circumstances. One possible reason could be conditionally defined macro calls. However, we do not know of any such case in the analysis and even if it would occur, we would still provide a generalized model, i.e., one with fewer constraints, but otherwise identical. For this reason, we expect that a better parsing algorithm has only a small impact on the overall accuracy. If this becomes an issue, it may be easily addressed as KernelHaven \cite{KernelHaven} supports the simple exchange of the used extractors.\looseness=-1

\textit{Completeness.} Section~\ref{sec:Further Information Sources} stresses the importance to consider all artifacts containing variability information. While we show exemplary how to incorporate variability information from non-code artifacts based on interface descriptions, we do not provide a complete integration of all artifact types at the moment. This is planned as future work.

\textbf{External validity.} This validity type addresses the generalizability of the obtained results to other contexts.

In the presented approach, a detailed knowledge about used variability realization techniques is required to provide a correct analysis of variability dependencies. This comprises knowledge of feature types, artifact types, and expressiveness of the parsed conditions. Already within the Bosch PS-EC product line, applying the approach to new products can be complicated. For instance, some products are not limited to pure relational expressions, as their subsystems use arithmetic expressions to check if certain bits of features are set. In general, this approach cannot be transferred to new product lines without detailed knowledge about their variability realization techniques.

\textbf{Construct validity} addresses the relation between the obtained result (the reverse engineered dependencies) and the real-world concept it reconstructs (the configuration domain knowledge). 

The feature effect approach detects for each feature the minimal precondition under which a value assignment impacts the product derivation. For this reason, we chose this approach to reverse engineer variability dependencies from the implementation, while ensuring that the outcome is no more restrictive than the implementation (cf.~\ref{obj:Lower Bound}). While these results are sufficient to filter irrelevant features in the configuration process, they do not represent the complete domain knowledge, e.g., partial configurations may be supported by the implementation but make no sense from a domain perspective. Thus, we plan to augment the extraction results with domain knowledge from the developers of the analyzed subsystems.

\vspace*{-12pt}
\section{Related Work}
\label{sec:Related Work}
In this paper, we presented an approach to reverse engineer variability dependencies from various artifact types. This work builds upon, but significantly extends prior work. Existing approaches of extracting variability models (features and their constraints), which are relevant to this work, can be divided into three categories according to their extraction strategy: mining of documented variability, analysis of variant instances, and variability model synthesis from annotated code. The remainder of this section is organized according to these three categories.

\textbf{Mining of documented variability.} Several studies provide concepts to extract feature models from variability documentation without analyzing code artifacts. Often old configurations are analyzed for this purpose. For instance, Lopez-Herrejon et al.\ \cite{Lopez-HerrejonGalindoBenavides+12} provide an evolutionary algorithm to compute a feature model based on feature sets (a kind of a configuration \cite{BenavidesSeguraRuiz-Cortes10}). This approach was tested on rather small models (up to 27 features) from S.P.L.O.T.\ \cite{MendoncaBrancoCowan09}, which differ fundamentally from realistic large-scale systems (several thousand features) \cite{BergerSheLotufo+13, SheLotufoBerger+10}. For the analysis of 896 configurations with 27 features, their algorithm requires already 11 minutes on a single core Intel Xeon E5620 VM. 
Czarnecki et al.\ \cite{CzarneckiSheWasowski08} present an algorithm to mine from old configurations a \textit{``Probabilistic Feature Model''}, which contains hard and soft constraints. Also, Davril et al.\ \cite{DavrilDelfosseHariri+13} use a similar approach, but they do not require a set of formal descriptions of product configurations. Instead, they utilize informal and potentially incomplete feature lists. While these approaches may deliver very good results, we were concerned about the generation of too restrictive results through the detection of false positive dependencies and, thus, violating \ref{obj:Lower Bound}. Nonetheless, mining old configurations provides an additional information source. For this reason, we currently use old configurations to verify our results.\looseness-1

Besides mining configurations further variability information sources can be used for the generation of a variability model. She et al.\ \cite{SheLotufoBerger+11} define a heuristic approach to identify feature hierarchies based on feature descriptions and propositional formulas. In later work, She et al.\ \cite{SheRysselAndersen14} present an algorithm for the creation of a feature model, encoding as much as possible dependencies in the tree structure rather than in textual cross-tree constraints. This approach requires only propositional formulas as input. Both papers focus on reverse engineering of feature structures out of propositional formulas, while our work focuses on reverse engineering of such formulas considering also numerical expressions.\looseness-1

\textbf{Analysis of variant instances.} This category addresses the extraction of configurable assets together with a variability model out of various product instances. These approaches assume that the analyzed product instances share some similar aspects, e.g., because they are developed via clone and own practices. For instance ExtractorPL \cite{ZiadiHenardPapadakis+14} and also its successor BUT4Reuse \cite{MartinezZiadiBissyande15} identify commonalities and variabilities in product instances and transfer them into configurable code together with a feature model for the configuration. This scenario differs from our scenario, as the Bosch PS-EC product line is developed as a configurable software product line.\looseness-1

Wille et al.\ \cite{WilleWehlingSeidl+17a} transfer concepts of reverse engineering variability of product lines to mine variability of enterprise architectures in order to provide recommendations for the reduction of unnecessary variability \cite{WehlingKennySeidl+17b}. This approach requires as input a structured set of technical architecture models for the identification of commonalities and variabilites. While the authors show that they are able to mine efficiently variability information from up to 1,200 architecture models, this approach would not be sufficient for the extraction of variability dependencies from various artifact types.

\textbf{Variability model synthesis from annotated code.} The most relevant work are techniques for reverse engineering of variability dependencies from implementation artifacts. Zhang and Becker \cite{ZhangBecker12, ZhangBecker13} project the structure of nested preprocessor statements directly to a variability tree, in which they map expressions that are syntactically identical to the same node. While they provide various complexity measures to analyze the generated variability tree, they do not provide methods for breaking down complex expressions to global conditions among the variables of the system. The same expression may also be mapped to different elements of the variability tree, if it occurs below different expressions in code artifacts.\looseness-1

Formal Concept Analysis (FCA) was originally developed to comprehend dependencies between complex preprocessed structures \cite{Snelting96}. Since then, this approach has been extended for the computation of feature models \cite{RysselPloennigsKabitzsch11, SheRysselAndersen14}. To reduce the number of concept latices, which is usually rather high, it is possible to consider constraints of the variability model if available or contradictory dependencies \cite{LuedemannAsadSchmid+16}. These papers focus on the synthesis of a feature model in a FODA-like fashion \cite{KangCohenHess+90}, rather than considering non-Boolean expressions and heterogeneous artifact types.

The feature effect approach developed by Nadi et al.\ \cite{NadiBergerKastner+14, NadiBergerKastner+15} provides a sound approach to abstract from the nesting of predecessor blocks and compound expressions to global dependencies between features of a variability model, even if this approach was only used for reverse engineering Boolean dependencies. The result are constraints among already existing features, without the need to create a new structure with possible artificial nodes. For this reason, we took this approach as basis for the variability analysis of the Bosch PS-EC product line. However, while our analysis take heterogeneous artifact types into account, Nadi et al.\ use a more accurate tooling to gather variability dependencies.

\section{Conclusion and Future Work}
\label{sec:Conclusion}
In this paper, we presented a study of reverse engineering variability dependencies in a long-lasting industrial product line of the Robert Bosch GmbH. While the underlying concept of feature effect analysis was already applied to publicly available case studies \cite{NadiBergerKastner+15}, this could not be transferred directly to this industrial product line due to the use of non-Boolean features and heterogeneous artifact types. This paper presents lessons learned and observations of reverse engineering variability in industrial context.

Section~\ref{sec:Product-wise Analysis} presents first results of the product-wise impact analysis of the reverse engineered variability dependencies. The relative amount of dependent features is between 31.8\% and 64.9\% when analyzing each product variant in isolation.
The dispersion of constrained features is broad, but comparable to the number of constrained features in variability models of publicly available product lines \cite{BergerSheLotufo+13}.
However, this amount decreases when the results of multiple products are combined. A preliminary cluster analysis indicates that this effect is probably related to the combination of products from different technical domains (gasoline vs.\ diesel) and can be cushioned by the development of separate variability models for these clusters. The results indicate a high potential to reduce the effort of configuring new product variants.

The presented approach successfully extracts variability dependencies of non-Boolean features from heterogeneous artifact types. We plan three extensions to the presented approach:

\textbf{Improved parsing of code files.} Currently, we use a block-based code extractor \cite{KH_CodeBlockExtractor} inspired by Undertaker \cite{Undertaker} as a lightweight and fast parser to extract presence conditions from code files. The presented tool chain allows to flexibly exchange the used parser. For the future, we plan to use TypeChef \cite{TypeChef} to parse code files, which provides more accurate results through macro-aware parsing and variability-aware type checking. However, this requires a complex configuration of TypeChef and drastically increases the run-time. It remains unclear whether an appropriate configuration of TypeChef is possible in order to handle adequately the huge amount of numerical expressions as used in the Bosch PS-EC product line. That is why we chose the block-based code extractor for setting up the analysis tool-chain.

\textbf{Mining old configurations.} For the Bosch PS-EC product line many old configurations are available. It is planned to use data mining to extract further variability dependencies.

\textbf{Configurations mismatch analysis.} The configuration mismatch analysis aims at identifying divergences between modeled and implemented dependencies \cite{El-SharkawyKrafczykSchmid17}. In this context, we can use either the consistency checks of code files (cf.\ Listing~\ref{lst:C-Example}) or manually developed feature models to verify the correctness of the reverse engineered variability. However, since the consistency checks may also be edited by developers it remains unclear whether an identified mismatch points to a false positive reported constraint or to an underspecified consistency check. Furthermore, the feature models specify the dependencies among features belonging to the same subsystem. Thus, they cannot be used to verify identified dependencies among features from different subsystems.

Finally, we plan a workshop to augment the extraction results with domain knowledge from the developers.

\begin{spacing}{0.9}
\begin{acks}
\vspace*{-1pt}
This work is partially supported by the ITEA3 project $\text{REVaMP}^2$, funded by the \grantsponsor{revamp}{BMBF (German Ministry of Research and Education)}{https://www.bmbf.de/} under grant \grantnum{revamp}{01IS16042H} and \grantnum{revamp}{01IS16042C} and extends work of the EvoLine project founded by the \grantsponsor{SPP1593}{DFG (German Research Foundation)}{http://www.dfg-spp1593.de/} under the Priority Programme \grantnum{SPP1593}{SPP 1593}. Any opinions expressed herein are solely by the authors and not by the BMBF or the DFG.
\end{acks}

\end{spacing}

\vspace*{-1pt}
\begin{spacing}{0.95}
\bibliographystyle{ACM-Reference-Format}
\bibliography{literature}
\end{spacing}

\end{document}